\documentclass[12pt]{article}
\usepackage{graphics}

\newcommand{\pp}{{p^{\prime}}}
\newcommand{\xbar}{\overline{x}}
\newcommand{\be}{\begin{equation}}
\newcommand{\ee}{\end{equation}}
\newcommand{\mat}[3]{\left<{#1}\left|{#2}\right|{#3}\right>}
\newcommand{\meas}[2]{\frac{d^{#2}{#1}}{(2\pi)^{#2}}}
\newcommand{\ra}{\rightarrow}
\newcommand{\MS}{\overline{\rm MS}}
\newcommand{\e}{\epsilon}
\newcommand{\prefac}[2]{\frac{{#1}\lambda^4
  }{{#2}(4\pi)^3 t^2} p_+^j}
\newcommand{\al}{\alpha}
\newcommand{\scale}{\frac{1}{\e} \left(\frac{4\pi\mu^2}{-t}\right)^\e}
\newcommand{\RR}{\mathcal{R}}
\newcommand{\mom}{\stackrel{\rm moments}{\longleftrightarrow}}

\newlength{\updownindent}
\newlength{\leftrightindent}
\begin{document}

\thispagestyle{empty}
\renewcommand{\thefootnote}{\alph{footnote}}

\begin{flushright}CERN-TH/99-90\\ ETH-TH/99-07\\ SWAT/99-223\\ \end{flushright}
\vspace{0.15in}

\begin{center}
  {\Large \bf Target Fragmentation in Semi-Inclusive DIS: Fracture Functions,
  Cut Vertices \& the OPE}\\
\vspace{0.3in}
  {\bf M. Grazzini}\\
\vspace{0.1in}
  {\it Institute for Theoretical Physics, ETH-H\"onggerberg,}\\
  {\it 8093 Z\"urich, Switzerland.}\\
  E-mail: {\tt grazzini@itp.phys.ethz.ch}\\
\vspace{0.2in}
  {\bf G. M. Shore}\\
\vspace{0.1in}
  {\it TH Division, CERN,} \\
  {\it CH-1211 Gen\`eve 23, Switzerland. }\\
    and\\
  {\it Physics Department, University of Wales Swansea,} \\
  {\it Singleton Park, Swansea SA2 8PP, UK.}\\
  E-mail: {\tt g.m.shore@swansea.ac.uk}\\
\vspace{0.2in}
  {\bf B. E. White}\\
\vspace{0.1in}
  {\it Physics Department, University of Wales Swansea,} \\
  {\it Singleton Park, Swansea SA2 8PP, UK.}\\
  E-mail: {\tt pywhite@python.swan.ac.uk}\\
\vspace{0.15in}
\end{center}

\begin{abstract}
We discuss semi-inclusive Deep Inelastic Scattering (DIS) in the
$z \ra 1$ limit, in particular the relationship between
fracture functions, generalised cut vertices and Green functions
of the composite operators arising in the OPE.
The implications, in the spin-polarised case, for testing
whether the ``proton spin'' effect is target-independent are explored.
Explicit calculations in $(\phi^3)_6$ theory are presented which are consistent
with our observations.
\end{abstract}

\vspace{0.3in}
\begin{flushleft}CERN-TH/99-90\\ ETH-TH/99-07\\ SWAT/99-223 -- March 1999\\
\end{flushleft}

\newpage
\setcounter{page}{1}

\renewcommand{\thefootnote}{\arabic{footnote}}
\setcounter{footnote}{0}

\section{Introduction}
\label{sec_intro}

The ``proton spin'' problem \cite{ESMC}
concerns the anomalous suppression of the
Ellis-Jaffe sum rule for the first
moment $\Gamma^p_1$ of the polarised proton structure function
$g_1$. In a series of papers, Narison, Shore and Veneziano~
\cite{SV2,NSV1,NSV2}
have shown that the underlying mechanism can be understood as
topological charge screening. The flavour singlet contribution to the sum-rule,
far from being a measure of quark or gluon spins, is in fact a measure of
gluon topological charge. The screening is a generic
property of the QCD vacuum rather
than a special property of the proton target.

This interpretation suggests that the ``proton spin'' suppression is in fact
a target-independent phenomenon, and recently \cite{SV1} it has been
proposed to test this directly in
polarised, semi-inclusive DIS, where we sum over
all possible final states containing a hadron $h$.
In the $z \ra 1$ limit, where $z$ is the fraction of the proton momentum
carried by $h$,
[fig.~\ref{fig_hadron}a] the polarised, semi-inclusive
structure function $\Delta M^{h/p}$ is modelled
by having the property of Regge factorisation into
a sum over products of p-h-$\RR$ couplings
and polarised structure functions of the Regge exchanges $g^\RR_1$.
By comparing experiments with different targets $N=p,n$ and different
tagged hadrons $h$, measurements of $\Delta M^{h/N}$ at $z\ra 1$ can
give ratios of the first moments of $g_1^\RR$ for reggeons $\RR$ with different
SU(3) flavour quantum numbers. According to the target-independence
conjecture, these ratios can be predicted simply in terms of SU(3)
group theoretic factors and a universal suppression factor determined by the
proton spin data.

A key requirement of the analysis of \cite{SV2,NSV1,NSV2}
was the identification
of the structure function with the matrix element of a local composite
operator arising in the OPE, and the subsequent decomposition of the
matrix element into a composite operator propagator and the
corresponding ``1PI'' vertex function. In order to provide a rigorous
theoretical foundation to the proposal of \cite{SV1} in the semi-inclusive
case, we need to show that a similar decomposition is valid.

In inclusive DIS, where we sum over all possible final hadronic states,
the structure functions factorise into hard, perturbatively calculable
factors and parton distribution functions,
which parametrise the non-perturbative physics:
\begin{center}\begin{tabbing}
  **\=Structure function***\= =***\=(Hard physics)*\=$\otimes$*\=\kill
  \>Structure function \>= \>(Hard physics) \>$\otimes$ \>
  (Parton dist. function).
\end{tabbing}\end{center}
The moments of the parton distributions can be represented either
as matrix elements of local operators or as space-like cut
vertices \cite{Mue,GM,BFK,Kub,Mun}. It has been shown explicitly
\cite{BFK,Mun} that these representations are equivalent, i.e.~in
the inclusive case, the cut vertices and local operator Green
functions are identical.

\begin{figure}
  \centering{\input{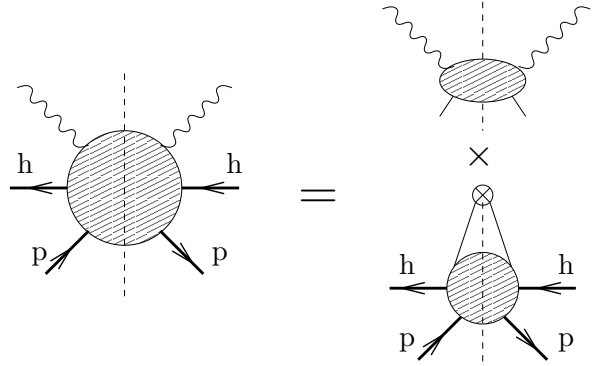}}
  \caption{Factorisation of the semi-inclusive structure functions. The
   blob on the bottom right is the fracture function.}
  \label{fig_fact}
\end{figure}

In the case of semi-inclusive DIS in the target fragmentation
region where $h$ has little transverse momentum relative to the
incoming proton, the semi-inclusive structure functions also
factorise into hard physics and {\it fracture} functions
\cite{TV,Grau,dFS,GTV,CGT}: [fig.~\ref{fig_fact}]
\begin{center}\begin{tabbing}
  **\=Semi-incl. struct. fn.***\= =***\=(Hard physics)*\=$\otimes$*\=\kill
  \>Semi-incl. struct. fn. \>=	\>(Hard physics)\>$\otimes$\>
  (Fracture function).
\end{tabbing}\end{center}
The fracture functions give the joint probability distribution for
finding a parton of specified type and a hadron $h$ in the proton.
Just as for the usual parton distributions, they absorb all the infra-red
singularities. Recently, it was shown \cite{GTV} that (extended)
fracture functions can be represented as generalised space-like cut
vertices. However, these generalised cut vertices cannot be identified
with local operator matrix elements; the familiar OPE analysis of inclusive
DIS is not directly applicable in the case of semi-inclusive DIS.

We can therefore summarise
the relationships between the various objects as follows:
\begin{center}\begin{tabbing}
  **\=Parton distributions***\= $\mom$***\=Cut vertices***\= =***\=\kill
  \>Parton distributions \>$\mom$ \> Cut vertices \>= \> Green functions \\
  \pushtabs
  **\=Fracture functions**\= $\mom$**\=
   Generalised cut vertices**\= $\neq$**\=\kill
  \>Fracture functions \>$\mom$ \>Generalised cut vertices \>$\neq$ \>
   Green functions \\
  \poptabs
\end{tabbing}\end{center}

At first sight, it therefore appears that the composite operator
propagator -- 1PI vertex methods of \cite{NSV1,NSV2} are not applicable
to semi-inclusive DIS. This is, however, only partially true.
The purpose of this paper is to examine precisely the relationship
between fracture functions, generalised cut vertices and Green functions,
and to determine under what circumstances the methods of \cite{NSV1,NSV2}
can be applied to semi-inclusive DIS. We will present evidence that
the generalised cut vertices simplify sufficiently in the $z \ra 1$
limit to allow the factorisation of the same composite operator propagator
from the semi-inclusive structure function as in the inclusive case.
This supports the proposal in ref.\cite{SV1} to test target independence
and universal topological charge screening in semi-inclusive DIS in the
$z \ra 1$ limit.

To make this assertion plausible, we have performed perturbative calculations
of generalised cut vertices to one loop in a simplified model expected
to share many features (e.g.~asymptotic freedom, collinear singularities, etc.)
of perturbative QCD, viz.~massless $\phi^3$ theory in six dimensions.
We find that, in the $z \ra 1$ limit, the dominant diagrams have such
a simple analytic structure that they can indeed be expressed as Green functions
in the same way as the inclusive case.

This paper is organised as follows: In section \ref{sec_proper}, we summarise
the ``proton spin'' effect; section \ref{sec_semi} provides definitions of
kinematic variables for semi-inclusive DIS and an overview of factorisation
of hard processes; our proof of the relation of generalised
cut vertices and Green functions is in section \ref{sec_cuts};
section \ref{sec_diags} contains
the results of the perturbative calculations
of the generalised cut vertices. Our conclusions and discussion of
the implications of these results are given in section \ref{sec_conc}.


\section{Local Operators and Proper Vertices}
\label{sec_proper}

In this section we summarise the resolution of the EMC/SMC
``proton spin'' problem proposed in refs.\cite{SV2,NSV1,NSV2,Sho}
in terms of gluon
topological charge screening.
We emphasise the need to express the target distribution functions
in terms of matrix elements of local operators.

The measurement of the first moment $\Gamma^p_1$ of the polarised
proton structure function by the EMC and SMC collaborations has shown
it to be suppressed relative to its OZI expectation. In other words,
$a^0$ is suppressed compared with $a^8$, where $a^0$ and $a^8$ appear in the
OPE of $\Gamma^p_1$:
\be
  \Gamma^p_1 = \int_0^1 dx\; g^p_1(x, Q^2) = \frac{1}{12}
  C^{\rm NS}_1(\alpha_s) \left(a^3 + \frac{1}{3}a^8\right)
  + \frac{1}{9}C^{\rm S}_1(\alpha_s) a^0(Q^2),
\ee
where
\be\begin{array}{c}\begin{array}{cc}
  \mat{p;s}{A^3_\mu}{p;s} = \frac{1}{2}a^3 s_\mu,
  & \phantom{*****}\mat{p;s}{A^8_\mu}{p;s} = \frac{1}{2\sqrt{3}}a^8 s_\mu,
  \end{array} \\
  \mat{p;s}{A^0_\mu}{p;s} = a^0(Q^2) s_\mu,
\label{mat}\end{array}\ee
with $s_\mu = \overline{u}_s(p)\gamma_\mu\gamma_5u_s(p)$ being
the proton polarisation vector. In the QCD parton model~\cite{AR,BFR}
\be\begin{array}{c}\begin{array}{cc}
  a^3 = \Delta u - \Delta d, &
  \phantom{*****}a^8 = \Delta u + \Delta d - 2\Delta s, \end{array} \\
  a^0(Q^2) = \Delta u + \Delta d + \Delta s - n_f \frac{\al_s}{2\pi}
  \Delta g(Q^2).
\end{array}\ee
$n_f$ is the number of quark flavours, while $\Delta u$, $\Delta d$,
$\Delta s$ and $\Delta g$ are the first moments of the polarisation
asymmetries of the parton distributions.

The OZI expectation is given by approximating
 $\Delta s \simeq 0 \simeq \Delta g$, so that $a^0 \simeq a^8$.
There are two important points: 
\begin{enumerate}
  \item The form factors $a^0$, $a^3$ and $a^8$ are (reduced)
matrix elements of local,
 composite operators.
  \item $a^0(Q^2)$ depends on the hard scale $Q^2$ through
the term proportional to $\Delta g(Q^2)$.
This is a result of the
chiral $U_A(1)$ anomaly.
This is already an indication that $a^0$ is not to be interpreted
as a measurement of spin in the QCD parton model \cite{Sho,Montp97}.
\end{enumerate}

One can explain the OZI violations in the following way. Because of the
$U_A(1)$ anomaly, the flavour singlet axial current $A_\mu^0$ obeys
an anomalous Ward identity written, neglecting quark masses,
\be
  \partial^\mu A_\mu^0 \;-\; 2n_f Q \;\simeq\; 0,
\label{ward}\ee
where $Q = \frac{\al_s}{8\pi}{\rm tr}G_{\mu\nu}\widetilde{G}^{\mu\nu}$
is the topological charge density. Using (\ref{ward}), one can express
\be
  a^0(Q^2) \;=\; \frac{1}{2M_p} 2n_f\mat{p}{Q}{p}.
\label{top}\ee
Thus, $a^0$ measures topological charge rather than parton spin. The
suppression of $a^0$ is due to topological charge screening by the
QCD vacuum. We can see this in terms of the external leg propagators
of the operator $Q$; the rhs. of (\ref{top}) can be decomposed into
zero-momentum propagators and 1PI vertices $\Gamma^{\rm 1PI}$, so that
\begin{eqnarray}
  a^0(Q^2) & \;=\; & \frac{1}{2M_p} 2n_f \left[
  \mat{0}{T(Q\,Q)}{0}\;\Gamma^{\rm 1PI}_{Q{\rm pp}}
  \;+\; \mat{0}{T(Q\,\Phi_5)}{0}\;\Gamma^{\rm 1PI}_{\Phi_5{\rm pp}}\right]
  \nonumber \\
  & = & \frac{1}{2M_p} 2n_f \left[
  \chi(0)\;\Gamma^{\rm 1PI}_{Q{\rm pp}}\;+\;
  \sqrt{\chi^\prime(0)}\;\Gamma^{\rm 1PI}_{\Phi_5{\rm pp}}\right],
\end{eqnarray}
where $\Phi_5$ is proportional \cite{SV2} to $\overline{q}\gamma_5q$ and
\be
  \chi(k^2) = i\int d^dx\; e^{ik\cdot x} \mat{0}{T(Q(x)Q(0))}{0}, \;\;\;\;\;\;
  \chi^\prime(0) = \left. \frac{d}{dk^2}\chi(k^2) \right|_{k^2 = 0}.
\ee

From the chiral Ward identities, it can be shown that $\chi(0)$ vanishes
for zero quark mass. Performing a similar decomposition
for $a^8$ results in a Goldberger-Treiman relation, in terms of
$\Gamma^{\rm 1PI}_{\Phi_5^8{\rm pp}}$,
for the flavour octet axial charge in the chiral limit.
With the assumption, justified in refs. \cite{SV2,NSV1,NSV2},
that the RG invariant
1PI vertices $\Gamma^{\rm 1PI}_{\Phi_5^8{\rm pp}}$ satisfy the OZI rule,
we find that
\be
  \frac{a^0(Q^2)}{a^8} = \left.\frac{\sqrt{6}}{f_\pi}\sqrt{\chi^\prime(0)}
  \right|_{Q^2}.
\ee
In this proposal, the suppression in $a^0(Q^2)$ is due to an anomalously
small value of the slope of the topological susceptibility $\chi^\prime(0)$
compared with its OZI expectation $f^2_{\pi}/6$.
$\chi^\prime(0)$ has been estimated using QCD spectral sum-rule techniques
\cite{NSV1,NSV2} and the results are in good
agreement with the measured suppression.\newline

\begin{figure}
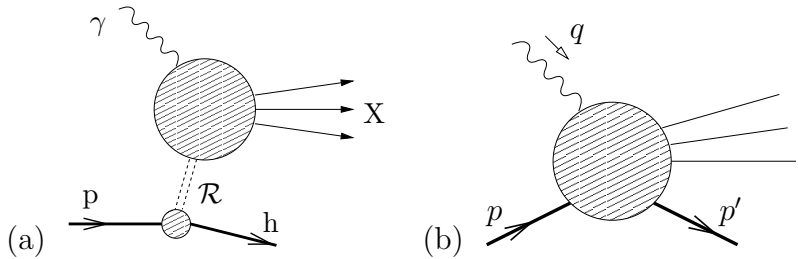

\begin{center}
\leavevmode
(a)\hspace{0.1in}{\input{diff_had.pstex_t}}\hspace{0.2in}
(b)\hspace{0.1in}{\input{hadron.pstex_t}}
\caption{(a) Amplitude in the $z \rightarrow 1$ limit. Proton $p$
  couples to hard hadron $h$ and reggeon $\RR$.
(b) Definitions of momenta for semi-inclusive DIS.}
\label{fig_hadron}
\end{center}
\end{figure}

We emphasise the following aspects of the analysis:
\begin{enumerate}
  \item The ratio $a^0/a^8$ does not depend on the target species; the
only references to ``proton'' were contained in the 1PI vertices
$\Gamma^{\rm 1PI}$.
  \item The procedure of extracting the propagators was dependent crucially
upon expressing $\Gamma^p_1$ in terms of matrix elements or Green functions
of local, composite operators.
\end{enumerate}
It is then natural to ask whether we can test this target-independence
conjecture directly. One possibility is to measure the polarised
structure functions of Regge poles
in the $-t/Q^2 \ll 1$, $z \rightarrow 1$
limit of polarised, semi-inclusive DIS.
[fig.~\ref{fig_hadron}a] We can model (cf. ref. \cite{BKS})
the polarised, semi-inclusive structure function for that process
$\Delta M^{h/p}(Q^2, \xbar, z, t)$ by
\be
  \Delta M^{h/p}(Q^2, \xbar, z, t)
  \stackrel{z \rightarrow 1}{\longrightarrow}
  F(t) (1 - z)^{-2\alpha_\RR(t)} g^\RR_1(Q^2, \xbar, t),
\ee
where $h$ is the hard, final state hadron and $g^\RR_1$ is the reggeon
$\RR$'s polarised structure function.
(The kinematic variables are defined in section
\ref{sec_semi}.)
By taking ratios of $\Delta M^{h/N}$ for different tagged hadrons $h$
and nucleon targets $N=p,n$, we can in effect measure ratios of the
$g_1^\RR$ for reggeons with different SU(3) flavour quantum numbers.
For example, if we choose $h$ to be a $\pi^-$, then there must be an
exchange with the the quantum numbers of a $\Delta^{++}$.
According to the target independence conjecture, these ratios will
depend only on SU(3) Clebsch-Gordon factors and a universal
suppression factor which can be extracted from the inclusive
``proton spin'' measurement.

In what follows, we will discuss whether such quantities as $g^\RR_1$
depend on the insertion of a local operator in the $z \ra 1$ limit,
as required if their flavour singlet first moments are indeed to be
governed by the slope of the topological susceptibility $\chi^\prime(0)$.

\section{Inclusive \& Semi-Inclusive DIS}
\label{sec_semi}

Here we define inclusive and semi-inclusive structure functions
for the simpler situation of scalar field theory, and we remind
the reader how the factorisation of the hard physics is
expressed.

We define the inclusive structure function as
\be
  W(q, p) = \frac{Q^2}{2\pi}\sum_X \int d^dx\;e^{iq\cdot x}
  \mat{p}{j(x)}{X} \mat{X}{j(0)}{p},
\ee
where $j(x) = \phi^2(x)$ plays the role of the electromagnetic current%
\footnote{
  There is no need to decompose $W$ into Lorentz scalar structure
functions since it already is a Lorentz scalar.
}
and the sum is over all possible final states.
$p$ and $q$ are the momenta of the ``proton'' and ``photon'' respectively.
Factorisation of the hard physics is written as
\be
  W(Q^2,x) = \int_x^1 \frac{du}{u}
  f(u) C(\frac{x}{u}, Q^2)
\ee
where $Q^2 = -q^2$ and $x = Q^2/2p\cdot q$.
$C$ is calculable perturbatively and $f$ is a ``parton'' distribution
function, parametrising the non-perturbative
physics. Taking moments wrt. $x$, we
have
\be
  \int^1_0 dx\; x^{j - 1} W(Q^2, x) = f^j C^j(Q^2),
\ee
where $f^j$ and $C^j$ are $j$th moments of $f$ and $C$.
$f^j$ can be represented either as a Green function of a composite operator
or as a space-like cut vertex. (Section \ref{sec_cuts}.)

To define the semi-inclusive structure function,
we constrain there to be a hadron $h$
of a specified type and momentum $\pp$ in the final state:
[fig.~\ref{fig_hadron}b]
\be
  W(q, p, \pp) = \frac{Q^2}{2\pi}\sum_X \int d^dx\;e^{iq\cdot x}
  \mat{p}{j(x)}{h,X} \mat{h,X}{j(0)}{p},
\ee
The convenient Lorentz-invariant variables for this process are as follows:
\be\begin{array}{cc}
  Q^2 = -q^2, & t = (p - \pp)^2, \\
  z = \frac{\pp\cdot q}{p\cdot q}, & x = \frac{Q^2}{2p\cdot q}.
\end{array}\ee
We also use
\be
\xbar = \frac{Q^2}{2(p - \pp)\cdot q} = \frac{x}{1 - z}.
\ee

There are two, distinct contributions to the process, distinguished by the
value of $t$, namely the target fragmentation region at small $t$
and the current fragmentation region at large $t$, so that
$W = W_{\rm curr} + W_{\rm targ}$. In the target
fragmentation region we may neglect quantities suppressed by
powers of $t/Q^2$.

The target fragmentation contributions can be handled by introducing
extended fracture functions $\mathcal{M}(x, z, t)$%
\footnote{
Note that extended fracture functions, which in the
literature are also referred to as diffractive parton densities
\cite{BS}, are distinct from fracture functions, which do not
depend on $t$. The latter are required when considering
$d^3\sigma/dQ^2dzdx$ rather than $d^4\sigma/dQ^2dtdzdx$. In that
case, one is not able to separate target and current fragmentation
regions; a telling symptom is that the renormalisation group
equation of an ordinary fracture function is not homogeneous
\cite{GTV,CGT}.
}
to parametrise the non-perturbative
physics, so that%
\footnote{ The justification for the factorisation of the hard
physics from the non-perturbative physics lies with
factorisation theorems which we do not discuss here \cite{BS,CS}.}
\be
  W_{\rm targ} (Q^2, \xbar, z, t) = \int_{\xbar}^1 \frac{du}{u}
  \mathcal{M}(u, z, t) C(\frac{\xbar}{u}, Q^2),
\label{targconv}\ee where $C$ are calculable perturbatively. It
has recently been shown that the moments of (extended) fracture
functions can be represented as generalised cut vertices
\cite{GTV}. (Section \ref{sec_cuts}.) This is the first step in the
chain of identifications described in the introduction amongst
fracture functions, cut vertices and Green functions, and is
central to our analysis.

Taking the moment with respect to $\xbar$ of
(\ref{targconv}) we get
\be
  \int^1_0 d\xbar\; \xbar^{j - 1}
  W_{\rm targ} (Q^2, \xbar, z, t) =
  \mathcal{M}^j(z, t) C^j(Q^2).
\label{gen_ope}\ee
where $\mathcal{M}^j$ and $C^j$ are $j$th moments of $\mathcal{M}$
and $C$ respectively.
(\ref{gen_ope}) generalises the leading twist part of the OPE to
semi-inclusive processes, with $\mathcal{M}^j$
given by the lowest twist generalised cut vertex.

Specialising further to the r\'{e}gime of interest, we consider
the $z \rightarrow 1$ limit. [fig.~\ref{fig_hadron}a] In the
context of QCD, this has been modelled with reggeon exchanges,
with some success in describing data \cite{BKS}. The asymptotic
behaviour of $\mathcal{M}$ would be
\be
  \mathcal{M}(\xbar, z, t) \simeq F(t) (1 - z)^{-2\alpha_R(t)} f_\RR(\xbar, t),
\label{regge}\ee where $f_\RR$ is interpreted as the
parton density of the reggeon.

The question whether factorisation of hard physics and Regge
factorisation, as expressed by (\ref{regge}), both be valid is not
clear \cite{BKS}. However, we find that in the analogous situation
in $(\phi^3)_6$ perturbation theory, the dominant 1-loop diagrams
as $z \ra 1$ have a similar form to that in eq. (\ref{regge}),
with a single propagator in place of the Regge pole.


\section{Cut Vertices}
\label{sec_cuts}

In this part, we define space-like cut vertices
\cite{Mue,GM,BFK,Kub,Mun} and show that they may be
identified with local operator matrix elements for inclusive DIS. We then
extend the argument to the semi-inclusive case, by defining generalised
space-like cut vertices, and deriving the conditions under which they can
be related to Green functions of local operators.

\begin{figure}
  \centering{\input{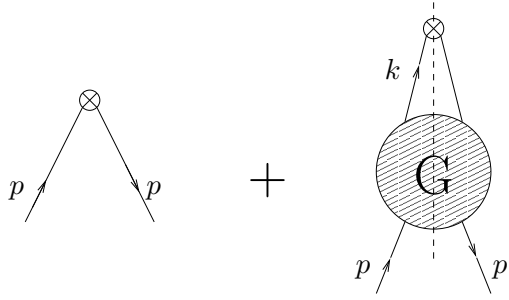}}
  \caption{Definition of cut vertex.}
  \label{fig_cut_vert}
\end{figure}

For the inclusive case, the lowest twist, spin $j$ cut vertex%
\footnote{
We require only the $+$ light-cone components of the derivatives,
since the rest are projected out by the Wilson co-efficients. Our conventions
are $x_\pm = \frac{1}{\sqrt{2}}(x_0 \pm x_3)$, $\partial_\pm x_\mp = 1$.
}
is: [fig.~\ref{fig_cut_vert}]
\be
  V_{\rm cut}(p) =
  (p_+)^j \;\; + \;\; \int \meas{k}{d} \Theta(0 < k_+ < p_+) (k_+)^j
  \stackrel{\rm \textstyle disc}{\scriptstyle (k - p)^2}G(k, p),
\label{cut_vert}\ee
where
\begin{eqnarray}
  \lefteqn{G(k,p) \; =} \nonumber \\
  & & \phantom{*} (\Delta_F(p))^{-2}
  \int d^dx d^dy d^dz \;\;e^{ip\cdot x - ik\cdot (y - z)}
  \mat{0}{T(\phi(x)\phi(y)\phi(z)
  \phi(0))}{0}.
\label{gamma2}\end{eqnarray}

The first term in (\ref{cut_vert}) is the
bare vertex, symbolised by $\otimes$, having two legs and Feynman rule
$(k_+)^j$, where $k_\mu$ is the momentum passing through it.
The second term is a convolution wrt. a loop momentum of the bare vertex
and a cut four-point function. The theta-function is encoded in the Feynman
rules by letting the broken line of the cut four-point function pass through
the $\otimes$.

\begin{figure}
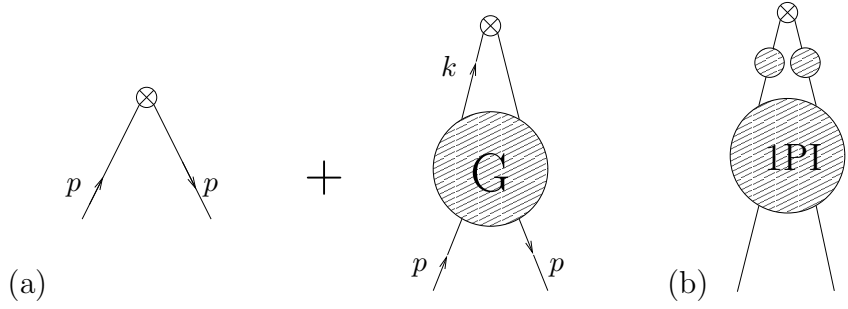

  \leavevmode
  \begin{center}
  (a)\hspace{0.1in}\input{uncut_vert.pstex_t}\hspace{0.5in}
  (b)\hspace{0.1in}\input{1PI.pstex_t}
  \end{center}
  \caption{(a) Definition of uncut vertex. (b) Decomposition of second
   term in (a) in terms of two-point functions and 1PI four-point function.}
  \label{fig_uncut_vert}
\end{figure}

\begin{figure}
  \centering{\input{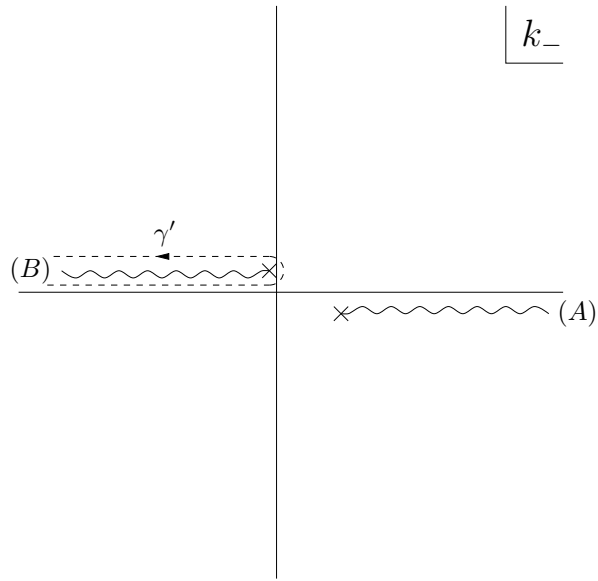}}
  \caption{Contour deformation when $0<k_+<p_+$.}
  \label{fig_cuts2}
\end{figure}

The cut vertex is identical to the corresponding matrix element,
[fig.~\ref{fig_uncut_vert}a] defined as
\begin{eqnarray}
  \lefteqn{V_{\rm uncut}(p)
  = \; \mat{p}{\: : \!\!\phi(-i\partial_+)^j\phi\!\! :(0)}{p} \; =}
  \nonumber \\
  & & \phantom{*}(\Delta_F(p))^{-2} \int \meas{k}{d} (k_+)^j
  \mat{0}{T(\phi(p)\phi(-k)\phi(k)\phi(-p))}{0}_{\rm disconn.}
\label{uncut_vert}\end{eqnarray}
where the ``disconn.'' label is an instruction to keep connected parts
and only those disconnected parts linked by the bare vertex. Thus,
\be
  V_{\rm uncut}(p)\; =\; (p_+)^j \;+ \;\Delta_F^{-2}(p)
  \int \meas{k}{d} (k_+)^j G(k, p).
\ee
To show that $V_{\rm uncut} = V_{\rm cut}$, we appeal to an argument of
Baulieu et al.: \cite{BFK} We note that $G(k, p)$ is not 1PI wrt. the
two legs carrying momentum $k$, and so has
three discontinuities depending on $k$, two in the variable $k^2$ and one
in $(k - p)^2$. In terms of 1PI Green functions, the uncut vertex is
as in [fig.~\ref{fig_uncut_vert}b].
We can parametrise these branch cuts individually by
\be
  (A)\;\; \left( \int_{s_0}^\infty d\mu(s)
  \frac{1}{k^2 - s + i\epsilon} \right)^2, \;\;\;\;\;\;\;\;
  (B)\;\; \int_{t_0}^\infty d\mu(t)
  \frac{1}{(k-p)^2 - t + i\epsilon}.
\ee
The branch cuts lie along
\be
  (A)\;\; k^2 = s - i\epsilon, \;\;\;\;\;\;\;\; (B)\;\; (k-p)^2 = t-i\epsilon,
\ee
with $s, t > s_0, t_0$ respectively. Now consider these regions in the
complex $k_-$ plane. Letting $k^2 = 2k_-k_+ - k^2_\perp$ etc. and choosing
$p_\perp = 0$, we get
\begin{figure}
  \centering{\input{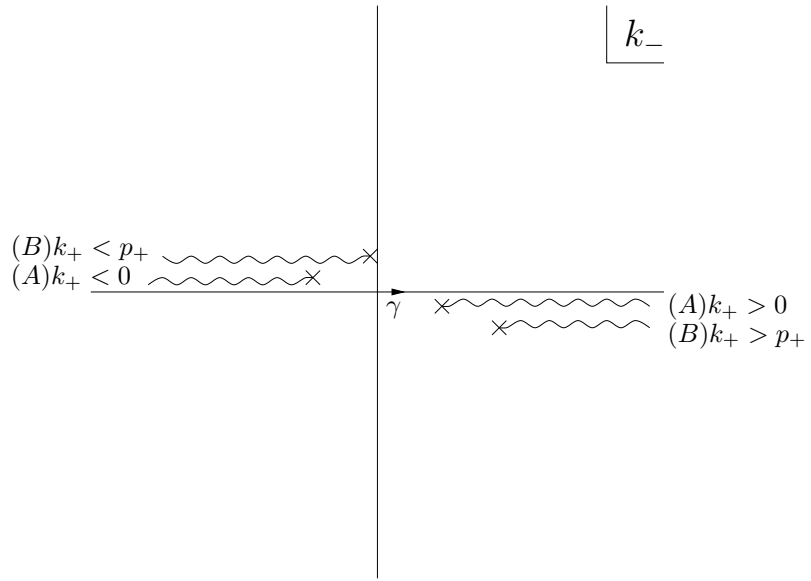}}
  \caption{Pole structure of uncut vertex in $k_-$ plane.}
  \label{fig_cuts1}
\end{figure}
\be
  (A)\;\; k_- = \frac{k^2_\perp + s - i\epsilon}{2k_+}, \;\;\;\;\;\;\;\;
  (B)\;\; k_- = p_- + \frac{k^2_\perp + t -i\epsilon}{2(k_+ - p_+)}.
\ee
The directions of the cuts depend on the sign of $k_+$ and
$(k - p)_+$. [fig.~\ref{fig_cuts1}] For $0 < k_+ < p_+$ and
$p_+ < k_+ < 0$, the branch cuts lie on opposite sides of the imaginary axis,
and there exists a contour deformation, $\gamma^\prime$, which wraps around
the branch cut (B) in [fig.~\ref{fig_cuts2}]. For the physical case of $p_+
> 0$, we only get a non-zero contribution from the discontinuity of $G$
wrt. $(k-p)^2$ in the interval $0 < k_+ < p_+$, as claimed.

Now we look at the semi-inclusive generalised cut vertices,
which give us the $j$th moments of
the fracture functions, defined at lowest twist, spin $j$
[fig.~\ref{fig_gen_cut_vert}] as
\begin{eqnarray}
  \lefteqn{V_{\rm cut}(p, \pp) \; = \; (p_+ - \pp_+)^j |\Lambda(p,\pp)|^2 \; +}
  \nonumber \\ & & \phantom{V_{\rm cut}}
  \int \meas{k}{d} \Theta(0 < k_+ < p_+ - \pp_+) (k_+)^j
  \stackrel{\rm \textstyle disc}{\scriptstyle (k - p + \pp)^2}G(k, p, \pp),
\label{gen_cut_vert}\end{eqnarray}
where
\be
  \Lambda(p,\pp) \; = \; \Delta_F^{-1}(p)\Delta_F^{-1}(\pp)\;
  \mat{0}{T(\phi(p)\phi(-\pp)\phi(-p + \pp))}{0},
\label{emm}\ee and \begin{eqnarray}
  \lefteqn{G(k,p,\pp) \; =} \nonumber \\
  & & \phantom{(k,p,\pp)} \Delta_F^{-2}(p)\Delta_F^{-2}(\pp)\;
  \mat{0}{T(\phi(p)\phi(-\pp)\phi(-k)\phi(k)\phi(\pp)\phi(-p))}{0}.
\end{eqnarray}
The first term in (\ref{gen_cut_vert}) replaces the bare vertex of the
previous case. We can also define an uncut vertex as follows:
\begin{figure}
  \centering{\input{gen_cut_vert.pstex_t}}
  \caption{Definition of generalised cut vertex.}
  \label{fig_gen_cut_vert}
\end{figure}
\begin{figure}
  \centering{\input{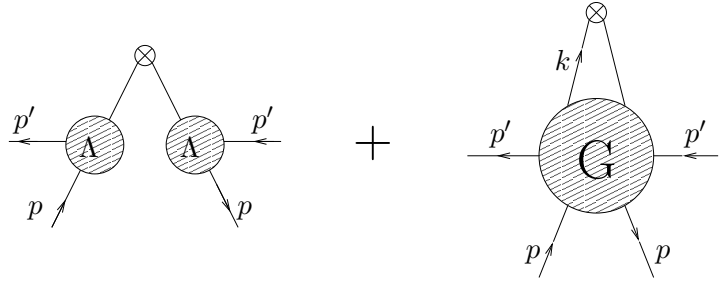}}
  \caption{Definition of generalised uncut vertex.}
  \label{fig_gen_uncut_vert}
\end{figure}
\be
  V_{\rm uncut}(p,\pp) \; = \; (p_+ - \pp_+)^j \Lambda^2(p,\pp) \;\;+\;\;
  \int \meas{k}{d} (k_+)^j G(k, p, \pp).
\label{gen_uncut_vert}\ee
$\Lambda^2 = |\Lambda|^2$, so the first terms in (\ref{gen_cut_vert})~\&~
(\ref{gen_uncut_vert}) are the same. The second terms are not the same,
however, because $G(k,p,\pp)$ has a more complicated pole structure;
it has branch cuts
in (A) $k^2$, (B) $(k-p)^2$, (C) $(k+\pp)^2$ and (D) $(k-p+\pp)^2$ and we
find it impossible to deform the $k_-$ contour around just one of these.
Putting $\pp_+ = zp_+ + O(t/Q^2)$, we find the pole structure as in
[fig.~\ref{fig_cuts3}]. If we restrict $p_+ > 0$ and $z > 0$, the contour
deformations are as follows: 
\pagebreak
\begin{figure}
  \centering{\input{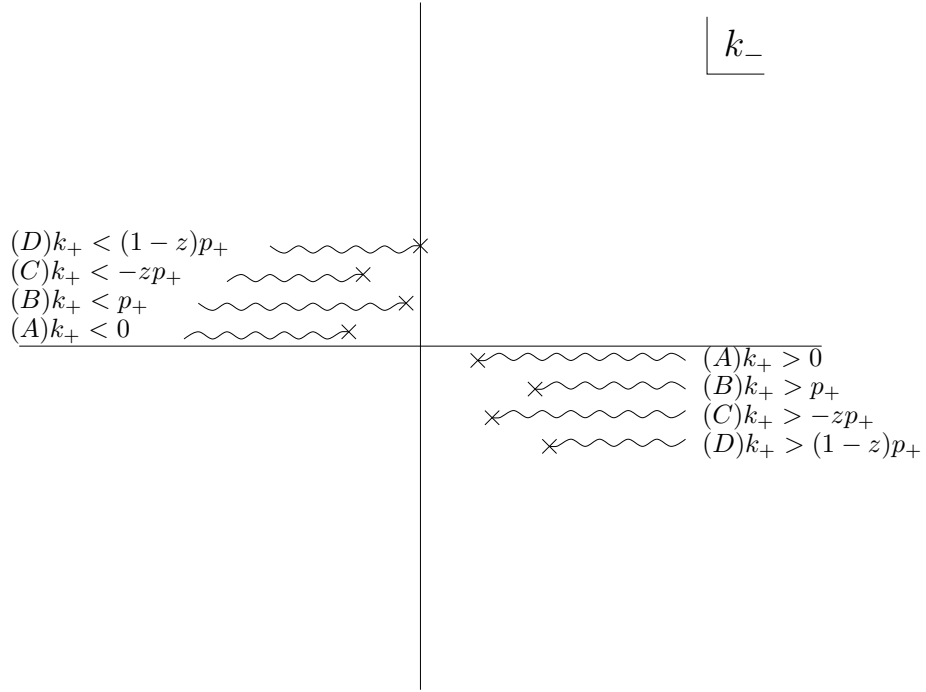}}
  \caption{Pole structure of generalised uncut vertex in $k_-$ plane.}
  \label{fig_cuts3}
\end{figure}
\begin{eqnarray}
 \lefteqn{\int \meas{k}{d} \left[
   \includegraphics[0in,0.3in][0.65in,0.85in]{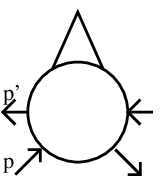}\right] \; =}
   \nonumber \\
    & & \int \meas{k}{d} \Theta(0<k_+<(1-z)p_+)
    \left[\includegraphics[0in,0.3in][0.65in,0.85in]{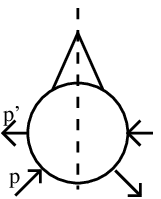}
    + \includegraphics[0in,0.3in][0.65in,0.85in]{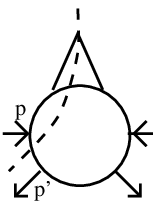}
    + \includegraphics[0in,0.3in][0.65in,0.85in]{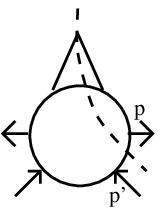}\right] \nonumber\\
    & \phantom{**}+\phantom{**} & \int \meas{k}{d} \Theta((1-z)p_+<k_+<1)
    \left[\includegraphics[0in,0.3in][0.65in,0.85in]{mini4.eps}
    + \includegraphics[0in,0.3in][0.65in,0.85in]{mini6.eps}\right] \nonumber\\
    & + & \int \meas{k}{d} \Theta(-zp_+<k_+<0)
    \left[\includegraphics[0in,0.3in][0.65in,0.85in]{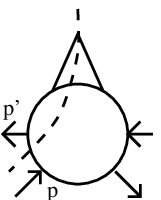}
    + \includegraphics[0in,0.3in][0.65in,0.85in]{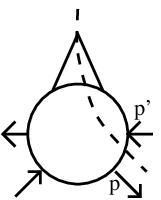}\right] \nonumber\\
 \lefteqn{= \; \int \meas{k}{d}
    \Theta\left( 0< \frac{k_+}{(1-z)p_+} <1 \right)
    \left[\includegraphics[0in,0.3in][0.65in,0.85in]{mini2.eps}\right]}
    \nonumber \\
    & + & \int \meas{k}{d}
    \Theta\left(-1 < \frac{k_+}{zp_+} < 0 \right)
    \left[\includegraphics[0in,0.3in][0.65in,0.85in]{mini3.eps}
    + \includegraphics[0in,0.3in][0.65in,0.85in]{mini5.eps}\right] \nonumber \\
    & + & \int \meas{k}{d} \Theta\left(0 < \frac{k_+}{p_+} <1 \right)
    \left[\includegraphics[0in,0.3in][0.65in,0.85in]{mini4.eps}
    + \includegraphics[0in,0.3in][0.65in,0.85in]{mini6.eps}\right]
\label{channels}
\end{eqnarray}

Eq. (\ref{channels}) is one of the main results of this paper.
Expressed generally, it is that the Green function of a (lowest twist)
composite operator is equal to the
sum of all possible cuts between the other external legs.

The first term in eq. (\ref{channels}) is the generalised cut vertex, with
the physical cut down the centre of the blob, but it
is joined by other contributions. Moreover, the theta-function
interval $0 < k_+ < (1-z)p_+$ vanishes as $z \ra 1$;
together with the factor $(k_+)^j$
in the integrand, this means that the physical cut vanishes at least as
fast as $(1-z)^j$.
However, there is a class of diagrams which do not have
the other cuts. These uncut diagrams can be identified with the physical cut
down the centre, as we will see in section
\ref{sec_diags}. For example, the Feynman diagram in
[fig.~\ref{fig_one_loop}f] can only be cut down the centre, giving a
contribution to the generalised cut vertex. This class turns out to contain
the dominant diagrams in the $z \ra 1$ limit.


\section{Explicit Calculations in $(\phi^3)_6$ Theory}
\label{sec_diags}

To see what happens to the perturbative expansion of the
generalised cut vertex as $z \ra 1$, we present the results of
calculations to one loop. These are in agreement with earlier
calculations \cite{Gra}, which compute the cut vertices
indirectly. We neglect contributions suppressed by powers of
$t/Q^2$. It is found that the limit $z \ra 1$ can be taken,
leaving as dominant contributions only those diagrams where the
$p$ and $\pp$ legs are connected to the rest of the diagram by
single propagators.

\begin{figure}
  \centering{\input{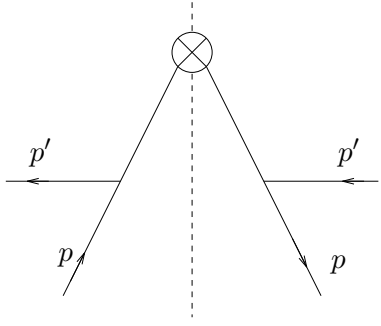}}
  \caption{Tree-level cut vertex}
  \label{fig_tree}
\end{figure}

To tree level, there is just one diagram, [fig.~\ref{fig_tree}] given by
\be
  V^{(0)}_{\rm cut} = \frac{\lambda^2}{t^2}p^j_+ (1-z)^j,
\ee
where $\lambda$ is the bare coupling constant. The factor $(1-z)^j$ appears
in all the diagrams and gives the leading order in $(1-z)$.

\begin{figure}
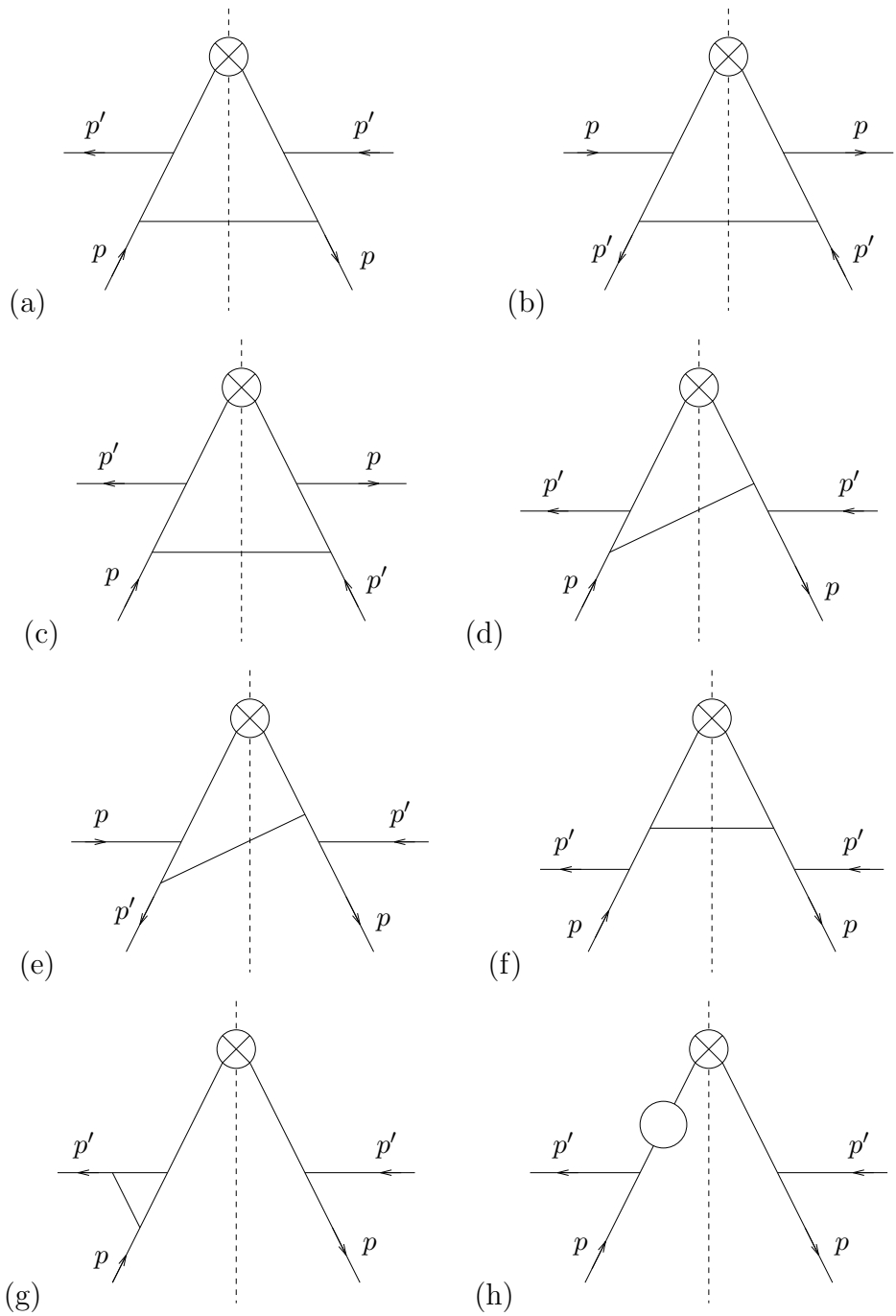

\leavevmode
\begin{center}
  (a)\hspace{0.1in}\input{diag1a.pstex_t}\hspace{0.5in}
  (b)\hspace{0.1in}\input{diag1b.pstex_t} \\ \vspace{0.1in}
  (c)\hspace{0.1in}\input{diag1c.pstex_t}\hspace{0.2in}
  (d)\hspace{0.1in}\input{diag1d.pstex_t} \\ \vspace{0.1in}
  (e)\hspace{0.1in}\input{diag1e.pstex_t}\hspace{0.2in}
  (f)\hspace{0.1in}\input{diag1f.pstex_t}  \\ \vspace{0.1in}
  (g)\hspace{0.1in}\input{diag1g.pstex_t}\hspace{0.2in}
  (h)\hspace{0.1in}\input{diag1h.pstex_t}
\caption{One-loop cut vertices}
\label{fig_one_loop}
\end{center}
\end{figure}

To one-loop order, there are eight diagrams [figs.~\ref{fig_one_loop}a-h]
and also their mirror images if they are not symmetric about the dotted line.
[Figs.~\ref{fig_one_loop}a\&b] are collinearly divergent, while
[figs.~\ref{fig_one_loop}f-h] are UV divergent. The rest are finite.
Calculating in the $\MS$ scheme, both types of divergence appear as simple
poles in $\epsilon$, where the space-time dimension is $d = 6 - 2\epsilon$.

The diagrams are evaluated by choosing a frame where, in light-cone
co-ordinates,
\be\begin{array}{l}
  p = (p_+, 0, 0_\perp),\\
  q = (-xp_+, q_-, 0_\perp),\\
  \pp = (\al^\prime p_+, \beta^\prime q_-, \pp_\perp), \\
  k = (\al p_+, \beta q_-, k_\perp),
\end{array}\ee
with
\be\begin{array}{l}
  \al^\prime = z + O(t/Q^2), \\
  \beta^\prime s = -t, \\
  \pp^2_\perp = -zt(1 + O(t/Q^2)), \\
  s = 2p\cdot q.
\end{array}\ee
Thus, for example, the diagram in [fig.~\ref{fig_one_loop}a] is
\begin{eqnarray}
  V^{(1a)}_{\rm cut} & \; =\; & \mu^{2\epsilon} \lambda^4 \int\meas{k}{d}
  \frac{2\pi\delta((k-p+\pp)^2)\;\Theta(0 < k_+ < p_+ - \pp_+)
  \;k^j_+}{k^4\;(k+\pp)^4} \nonumber \\
  & \; =\; & \frac{\mu^{2\epsilon}
  \lambda^4 p^j_+s}{4\pi} \int_0^{1-z} d\al\;\al^j
  \int d\beta \int\meas{k_\perp}{d-2} \nonumber \\
  & & \phantom{**} \times\frac{
  \delta((\al+z-1)(\beta s - t) - (k_\perp + \pp_\perp)^2)}{
  (\al\beta s - k^2_\perp)^2 \;
  ((\al + z)(\beta s - t) - (k_\perp + \pp_\perp)^2)^2} \nonumber \\
  & \; =\;  & \prefac{}{}(1-z)^{j+2} \int_0^1 \frac{d\al\;\al^j(1-\al)}{
  (\al+z(1-\al))^2} \; \left[-\scale\Gamma(2+\epsilon) \right. \nonumber \\
  & & \left. \phantom{\frac{!}{!}**} + \; 2\ln(\al+z(1-\al))
  \; +\; \ln(1-\al)\; -\; \ln\al\; -\; 1 \right].
\end{eqnarray}
Likewise, the other diagrams give, multiplying non-symmetric diagrams by 2,
\begin{eqnarray}
  V^{(1b)}_{\rm cut} & \; =\; &
  \prefac{}{}(1-z)^{j+2} \int_0^1\frac{d\al\;\al^j
  (1-\al)}{(1-\al(1-z))^2} \; \left[-\scale\Gamma(2+\epsilon)
  \right. \nonumber \\
  & & \left. \phantom{\frac{!}{!}} -\; 2\ln z\;  +\; \ln(1-\al)
  \; +\; 2\ln(1-\al(1-z)) \; -\; \ln\al\; -\; 1 \right], \nonumber \\
\end{eqnarray}
\begin{eqnarray}
  V^{(1c)}_{\rm cut} & \; =\; & 2\times\prefac{}{2}\frac{(1-z)^{j+2}}{z}
  \int_0^1 d\al\;\al^j(1-\al) \left[ \phantom{\frac{!}{!}} 4\ln(1-z) \right.
  \nonumber \\
  & & \phantom{\frac{!}{!}}\left.
  + \ln\left(\frac{\al(1-\al)}{(1-(1-z)(1-\al))^2}\right)
   + \ln\left(\frac{\al(1-\al)}{(z+(1-z)(1-\al))^2}\right)\right],
  \nonumber \\
\end{eqnarray}
\begin{eqnarray}
  V^{(1d)}_{\rm cut} & \; =\; & 2\times\prefac{}{}\frac{(1-z)^{j+1}}{z}
  \int_0^1 d\al\;\al^j[\ln(\al+z(1-\al))\; -\; \ln\al],
\end{eqnarray}
\begin{eqnarray}
  V^{(1e)}_{\rm cut} & \; =\; & 2\times\prefac{}{}(1-z)^{j+1}
  \int_0^1 d\al\;\al^j[\ln z + \ln\al - \ln(1-\al(1-z))],
  \nonumber \\
\end{eqnarray}
\begin{eqnarray}
  V^{(1f)}_{\rm cut} & \; =\; &
  \prefac{}{}(1-z)^j\scale\int_0^1 d\al\;\al^j(1-\al).
\end{eqnarray}
The rest are straightforward virtual loop corrections giving
\begin{eqnarray}
  V^{(1g)}_{\rm cut} & \; =\; & 2\times\prefac{}{2}(1-z)^j\scale,
\end{eqnarray}
\begin{eqnarray}
  V^{(1h)}_{\rm cut} & \; =\;  & 2\times\; -\prefac{}{12}(1-z)^j\scale.
  \phantom{***************************************}
\end{eqnarray}
Taking the limit $z\ra 1$ and, keeping only leading orders in $(1-z)$,
we have
\begin{eqnarray}
  V^{(1a)}_{\rm cut} & \; =\; & \prefac{}{}(1-z)^{j+2}\frac{1}{(j+1)(j+2)}
  \; \left[-\scale\Gamma(2+\epsilon) \right. \nonumber \\
  & & \left. \phantom{\frac{!}{!}***}
  -\; \gamma_E\; -\; \psi(j+1) \right]\; + \; O((1-z)^{j+3}),
\end{eqnarray}
\begin{eqnarray}
  V^{(1b)}_{\rm cut} & \; =\; & V^{(1a)}_{\rm cut}\;+\;O((1-z)^{j+3}),
\end{eqnarray}
\begin{eqnarray}
  V^{(1c)}_{\rm cut} & \; =\; & 2\times\; -\prefac{}{}(1-z)^{j+2}
  \frac{1}{(j+1)(j+2)}\nonumber \\
  & & \phantom{***} \times\left[ 2\ln\left(\frac{1}{1-z}\right)
  \; -\; 1\; +\; \gamma_E + 2\psi(j+3)-\psi(j+1) \right] \nonumber \\
  & + & O\left(\ln\left(\frac{1}{1-z}\right)(1-z)^{j+3}\right),
\end{eqnarray}
\begin{eqnarray}
  V^{(1d)}_{\rm cut} & \;=\; & 2\times\;
  \prefac{}{}(1-z)^{j+1}\frac{1}{(j+1)^2} \;+\; O((1-z)^{j+2}),
\end{eqnarray}
\begin{eqnarray}
  V^{(1e)}_{\rm cut} & \;=\; &
  2\times\; -\prefac{}{}(1-z)^{j+1}\frac{1}{(j+1)^2}
  \;+\; O((1-z)^{j+2}),
\end{eqnarray}
\begin{eqnarray}
  V^{(1f)}_{\rm cut} &\;=\; & \prefac{}{}(1-z)^j\scale\frac{1}{(j+1)(j+2)},
\end{eqnarray}
\begin{eqnarray}
  V^{(1g)}_{\rm cut} &\;=\; & 2\times\prefac{}{2}(1-z)^j\scale,
\end{eqnarray}
\begin{eqnarray}
  V^{(1h)}_{\rm cut} & \;=\; & 2\times\; -\prefac{}{12}(1-z)^j\scale.
\end{eqnarray}

In the above, finite parts proportional to $p^j_+(1-z)^j/t^2$ are neglected.
To renormalise the UV poles of [figs.~\ref{fig_one_loop}f-h], we introduce
the running coupling constant $\lambda(t)$ in terms of the bare coupling
$\lambda$,
\be
  \lambda^2 = \lambda^2(t)\left(1 - \frac{3}{4}\frac{\lambda^2}{(4\pi)^3}
  \scale\right),
\ee
and multiply by a renormalisation factor $Z^{-2}_\phi Z_j$, $Z_j$ being for
the spin $j$ bare vertex, such that%
\footnote{
Here, the renormalisation prescription differs from that of ref.~\cite{Gra},
which simply introduces the running coupling.
}
\be
  Z_\phi \;=\; 1 + \frac{1}{12}\frac{1}{\epsilon}\frac{\lambda^2}{(4\pi)^3}
\ee
and
\be
  Z_j \;=\; 1 - P_j\frac{1}{\epsilon}\frac{\lambda^2}{(4\pi)^3},
\ee
where $P_j$ is the $j$th moment of the AP kernel for $(\phi^3)_6$ theory,
$P(x) = x(1-x) - \frac{1}{12}\delta(1-x)$:
\be
  P_j = \int_0^1 \frac{dx}{x}x^j P(x) = \frac{1}{(j+1)(j+2)} - \frac{1}{12}.
\ee

The final result, to order $(1-z)^j$, is
\[
  V^{\rm 1-loop}_{\rm cut, ren.}(p,\pp) \;=\; \frac{\lambda^2(t)}{t^2}
  p^j_+(1-z)^j \left[1 + \frac{\lambda^2}{(4\pi)^3}
  \ln\left(\frac{4\pi\mu^2}{-t}\right)\left(\frac{1}{6} + P_j\right)\right]
\]\be
  +\; O\left(\ln\left(\frac{1}{1-z}\right)(1-z)^{j+2}\right).
\ee

The observation we make is that only diagrams in [figs.~
\ref{fig_tree}\&\ref{fig_one_loop}f-h] contribute to leading order
in $(1-z)$.

\begin{figure}
  \centering{\input{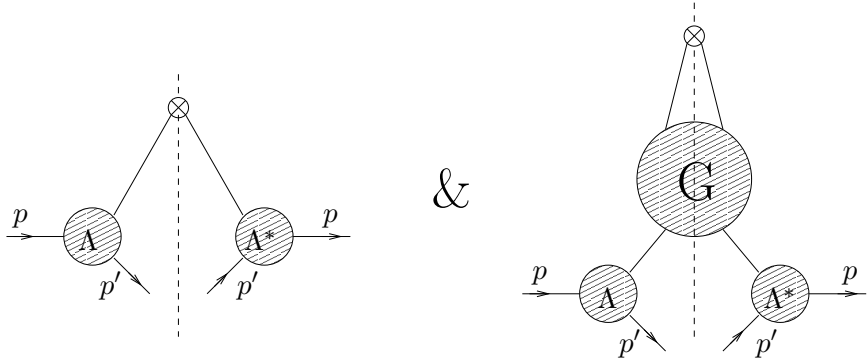}}
  \caption{Topology of the dominant one-loop diagrams.}
  \label{fig_dom_cuts}
\end{figure}

The diagrams in [figs.~\ref{fig_tree}\&\ref{fig_one_loop}f-h]
have the topology of
[fig.~\ref{fig_dom_cuts}], expressible as
\begin{eqnarray}
  \lefteqn{V_{\rm cut}(p,\pp) \stackrel{z \ra 1}{\longrightarrow}
  \left[ (p_+ - \pp_+)^j \;+\; \phantom{\meas{k}{d}}
  \right.} \nonumber \\
  & & \left.
  \int\meas{k}{d}\;\Theta(0 < k_+ < p_+-\pp_+)\;
  (k_+)^j\;\stackrel{\rm \textstyle disc}{\scriptstyle (k - p+\pp)^2}
  G(k,p-\pp)\right] |\Lambda(p,\pp)|^2, \nonumber \\
\label{dom_cuts}\end{eqnarray} with $G$ and $\Lambda$ as defined in
(\ref{gamma2})\&(\ref{emm}). Note that $G$ is collinear-safe to
one-loop since the lines carrying $p-\pp$ are off shell by $t$.
Now, recall that in section \ref{sec_cuts} it was shown that uncut
diagrams were equal to a sum of all possible cuts between external
legs. For the dominant diagrams, the only cuts are the ones
contributing to $V_{\rm cut}(p,\pp)$; eq. (\ref{dom_cuts}) may be
equated with its uncut analogue:
\be
  \left[ (p_+ - \pp_+)^j \;+\; \int\meas{k}{d}\;(k_+)^j\;G(k,p-\pp)\right]
  \Lambda^2(p,\pp)
\ee
Thus, we find that at small $(1-z)$ the generalised cut vertices depend
approximately on the Green function of a composite operator.


\section{Discussion \& Conclusions}
\label{sec_conc}

The main conclusion of sections \ref{sec_cuts}-\ref{sec_diags} is the
following: If the dominating contribution to the fracture function
has the topology of [fig.~\ref{fig_dom_cuts}], then the fracture function
can be identified with an uncut amplitude. Now we discuss whether this
is the case for QCD.

Although perturbative calculations are not appropriate to describe the process
in question, they can give insight into certain features of the
process which can be expected to be independent of non-perturbative effects.
The dominance of a particular class of processes as $z \ra 1$ is
one of them. First, we give a simple
argument to explain the $(1-z)$ dominance of this class of process in the
perturbative expansion, and compare it with
another perturbative model which has been
studied by Hautmann et al. \cite{HKS}. Secondly, we discuss
the Regge hypothesis of ref.~\cite{SV1}.

\begin{figure}
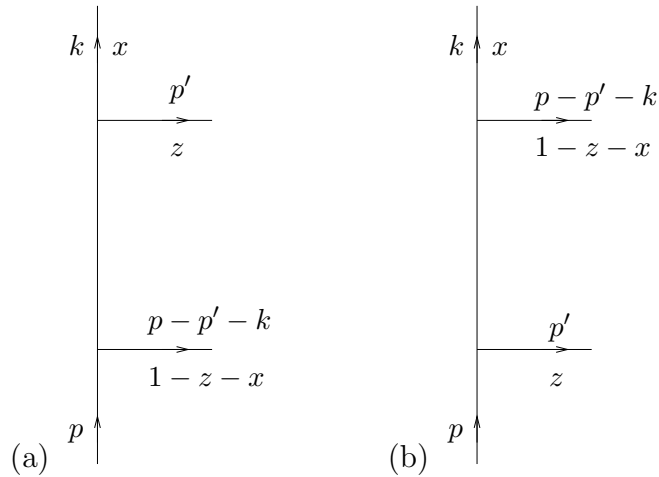

\begin{center}
\leavevmode
(a)\hspace{0.1in}\input{brancha.pstex_t}\hspace{0.8in}
(b)\hspace{0.1in}\input{branchb.pstex_t}
\caption{Matrix elements corresponding to
  (a) fig. \ref{fig_one_loop}a \&
  (b) fig. \ref{fig_one_loop}f. The external legs are marked with
    4-momenta and light-cone components scaled by $p_+$, such that $k_+ = xp_+$
    and $\pp_+ = z p_+$.}
\label{fig_branch}
\end{center}
\end{figure}

Let us contrast the behaviour of the diagrams in
[figs.~\ref{fig_one_loop}a\&f], the first of which is suppressed in $(1-z)$.
These have matrix elements as in
[figs.~\ref{fig_branch}a\&b]. In (b), the propagator connecting the $\pp$ leg
and the $(p-\pp-k)$ leg is fixed and off shell by $t$. In (a), the propagator
is constrained by the condition that the $(p-\pp-k)$ leg be on shell and equals
\be
  \frac{1}{(k + \pp)^2} = - \frac{(1-x-z)}{(k_\perp + \pp_\perp)^2}
  \sim \frac{1-z}{t},
\ee
at small $k_\perp$. Diagrams in [fig.~\ref{fig_one_loop}a-c] have two such
propagators off shell by $\frac{t}{1-z}$ and are suppressed by $(1-z)^2$ as
a result, while [fig.~\ref{fig_one_loop}d-e] have one, and go like $(1-z)$
instead.

\begin{figure}
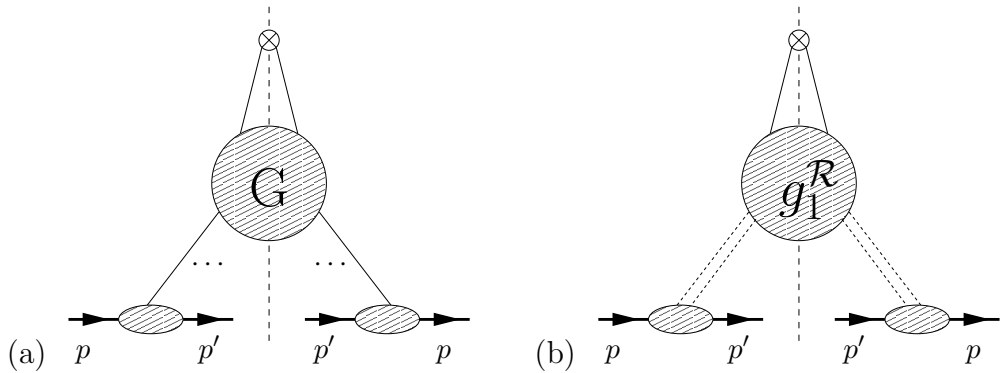

  \begin{center}
  (a)\hspace{0.1in}\input{diff_vert.pstex_t}\hspace{0.2in}
  (b)\hspace{0.1in}\input{regge_vert.pstex_t}
  \end{center}
  \caption{(a) Virtual exchanges between the bare vertex
   and the hadronic states in the $z \ra 1$ limit. (b) $z \ra 1$
   limit of the polarised fracture function, with $g^\RR_1$ the
   polarised reggeon structure function.}
  \label{fig_diff_vert}
\end{figure}

The rule appears to be that if the line carrying the momentum component
$zp_+$ branches out with soft emissions into the final state, its contribution
is small. If this carries over into the non-perturbative physics, then the
dominant class of processes
is as in [fig.~\ref{fig_diff_vert}a]. Note:
The factor $G$ is a {\it connected}, cut amplitude.
If we tried
to fit, say, [fig.~\ref{fig_one_loop}a] into this topology, it would
give a disconnected contribution to $G$, since there is a soft line
(carrying $(p-\pp-k)$) going into the final state.

In a perturbative QCD heavy-quark model for such a process, Hautmann
et al. \cite{HKS}
also find that such diagrams are dominant, with the single propagators
replaced by gluon pairs in a colour singlet. In that case, $G$ can be
defined in terms of cut Green functions, although these are not then equal
to uncut ones: $G$ is connected to the hadronic states by more than one
virtual line. However, this is only a difficulty in QCD perturbation
theory because of the multiparticle poles.
The Regge model proposed in ref.~\cite{SV1} provides a non-perturbative
description of the virtual exchanges [fig. \ref{fig_diff_vert}b].
Here, we expect the pole structure to be simple enough
to uncut the amplitude in the same
way as the $\phi^3$ diagrams: The reggeon structure function $g^\RR_1$ is equal
to a connected 3-point amplitude depending on an insertion of a
local operator. It is possible to perform
the decomposition of this amplitude into 1PI vertices and two-point
functions, as outlined in section \ref{sec_proper}.

To summarise, we have shown that in the limit $z \ra 1$, fracture functions
are dominated by a special class of process. If a Regge model for this
is realistic, these processes are related to an uncut amplitude.
This would mean that that the polarised reggeon structure function
can be used as a testing-ground for the target-independence hypothesis
of the ``proton spin effect''.


\subsection*{Acknowledgements}
We should like to thank Z.~Kunszt, L.~Trentadue
and G.~Veneziano for discussions during the course of this research.
The work of M.G. is supported by the
Swiss National Foundation. G.M.S. and B.E.W. are supported by
PPARC grant GR/L56374 in the UK. G.M.S. acknowledges EC TMR Network Grant
FMRX-CT96-0008. B.E.W. also wishes to thank PPARC for a
research studentship. 


\end{document}